\documentclass[10pt]{article}
\usepackage{graphicx}

\usepackage{amsmath}
\usepackage{epsfig}
\usepackage[T1]{fontenc} 
\usepackage{longtable}
\usepackage{psfrag} 

\def\Title#1{\begin{center} {\Large #1 } \end{center}}
\def\Author#1{\begin{center}{ \sc #1} \end{center}}
\def\Address#1{\begin{center}{ \it #1} \end{center}}

\newcommand\pubblock{\rightline{\begin{tabular}{l} Proceedings of the Second Annual LHCP\\ \pubnumber\\
         \pubdate  \end{tabular}}}

\newenvironment{Abstract}{\begin{quotation} \begin{center}
             \large ABSTRACT \end{center}\bigskip
      \begin{center}\begin{large}}{\end{large}\end{center} \end{quotation}}

\newenvironment{Presented}{\begin{quotation} \begin{center}
             PRESENTED AT\end{center}\bigskip
      \begin{center}\begin{large}}{\end{large}\end{center} \end{quotation}}

\def\beq{\begin{equation}}
\def\eeq#1{\label{#1}\end{equation}}
\def\eeqn{\end{equation}}

\def\beqa{\begin{eqnarray}}
\def\eeqa#1{\label{#1}\end{eqnarray}}
\def\eeqan{\end{eqnarray}}

\let\bar=\overbar

\def\Dslash{\not{\hbox{\kern-4pt $D$}}}
\def\dslash{\not{\hbox{\kern-2pt $\del$}}}

\def\msb{{\bar{\ssstyle M \kern -1pt S}}}

\usepackage{color}

\newcommand\pubnumber{ IFT-UAM/CSIC-14-091; FTUAM-14-35}

\newcommand\pubdate{\today}

\def\Comment#1{}
\newcommand{\bean}{\begin{eqnarray*}}
\newcommand{\eean}{\end{eqnarray*}}

\newcommand{\gapproxeq}{\lower
.7ex\hbox{$\;\stackrel{\textstyle >}{\sim}\;$}}
\newcommand{\lapproxeq}{\lower
.7ex\hbox{$\;\stackrel{\textstyle <}{\sim}\;$}}

\newcommand\lsim{\mathrel{\rlap{\lower4pt\hbox{\hskip1pt$\sim$}}
    \raise1pt\hbox{$<$}}}
\newcommand\gsim{\mathrel{\rlap{\lower4pt\hbox{\hskip1pt$\sim$}}
    \raise1pt\hbox{$>$}}}
\newcommand{\ba}{\begin{array}}
\newcommand{\ea}{\end{array}}
\newcommand{\nn}{\nonumber}

\newcommand{\be}{\begin{equation}}
\newcommand{\eequ}{\end{equation}}
\newcommand{\bear}{\begin{eqnarray}}
\newcommand{\eear}{\end{eqnarray}}

\newcommand{\cO}{{\cal O}}

\newcommand{\mL}{\mathcal{L}}

\newcommand{\mF}{\mathcal{F}}

\newcommand{\mM}{\mathcal{M}}

\newcommand{\Frac}[2]{\frac{\displaystyle #1}{\displaystyle #2}}

\begin{document}

\large
\begin{titlepage}
\pubblock

\vfill
\Title{  Electroweak Chiral Lagrangians and
$\gamma\gamma\to  Z_L Z_L,\, W_L^+ W_L^-$ at One Loop}
\vfill


\Author{R.L. Delgado, A. Dobado}
\Address{
Departamento de F\'isica Te\'orica I, UCM, \\
Universidad Complutense de Madrid,
  Avda. Complutense s/n,
28040 Madrid, Spain}

\Author{M.J. Herrero and  J.J. Sanz-Cillero~\footnote{Speaker.\\{}\\
JJSC would like to thank J.~Hollar, M.~Luszczak, M.A.~Pleier and P.~Tan
for useful discussions and advices
during the LHCP14 Conference and the organizers for their excellent work.
He would also like to thank BNL, JLab and UMass for their hospitality.
This work has been partly supported by
the European Union FP7 ITN INVISIBLES (Marie Curie Actions, PITN- GA-2011- 289442),
by the CICYT through the projects FPA2012-31880, FPA2010-17747,    CSD2007-00042, 
FPA2011-27853-C02-01 
and FPA2013-44773-P,   
by the CM (Comunidad Autonoma de Madrid) through the project HEPHACOS S2009/ESP-1473,
by the Spanish Consolider-Ingenio 2010 Programme CPAN (CSD2007-00042)
and by the Spanish MINECO's grant BES-2012-056054
and "Centro de Excelencia Severo Ochoa" Programme
under grant SEV-2012-0249.}}
\Address{
Departamento de F\'isica Te\'orica and Instituto de F\'isica Te\'orica,
IFT-UAM/CSIC\\
Universidad Aut\'onoma de Madrid,  C/ Nicol\'as Cabrera 13-15, \\
Cantoblanco, 28049 Madrid, Spain}


\vspace*{.7cm}

\vfill
\begin{Abstract}
In these proceedings we discuss our recent work
on $\gamma\gamma\to W^+_L W^-_L$ and $\gamma\gamma\to  Z_L Z_L$
within the framework of Electroweak Chiral Lagrangians with a light Higgs.
These observables are good candidates to provide indications of physics
beyond the Standard Model and can complement other analyses and global fits.
Making use of the equivalence theorem,
we have performed the computation up to the next-to-leading order in the chiral expansion,
i.e., including one-loop corrections and the full set of possible counterterms allowed
at that order in the low-energy effective field theory.
The one-loop amplitudes turn out to have a extremely simple structure
and they are ultraviolet finite. Thus, the relevant combinations of
higher order chiral couplings $c_\gamma^r$ and $(a_1^r-a_2^r+a_3^r)$ do not run.
We also identify an additional set of observables that depend on these same parameters,
which can be use to complement the $\gamma\gamma$ analysis through a global fit.
%

\end{Abstract}
\vfill

\begin{Presented}
The Second Annual Conference\\
 on Large Hadron Collider Physics \\
Columbia University, New York, U.S.A \\
June 2-7, 2014
\end{Presented}
\vfill
\end{titlepage}
\def\thefootnote{\fnsymbol{footnote}}
\setcounter{footnote}{0}
%

\normalsize


\section{Introduction}

Based on the article~\cite{photon-scat}, we discuss in these proceedings
our study of the photon-photon scattering into longitudinal weak gauge bosons,
$\gamma\gamma\to Z_L Z_L, W_L^+ W_L^-$.
This observable can be an optimal place
to look for deviations from the Standard Model (SM) in future analyses
and to test the composite nature of the Higgs.
Since this boson does not contribute at tree-level to the photon-photon scattering in the SM, any new physics in the Higgs is expected
to give particular signatures. In combination with other observables it is possible to extract the relevant couplings of the low-energy
effective field theory (EFT), the Electroweak Chiral Lagrangian
including the Higgs (ECLh)~\cite{photon-scat}.

On the experimental side, there are dedicated programs at LHC on photon-photon scattering.
The CMS Collaboration has published the first results on
$\gamma\gamma\to W^+ W^-$~\cite{photon-scat-CMS},
showing the feasibility of this type of analysis.
There are indeed the approved projects  CMS-TOTEM~\cite{CMS-TOTEM}
and  ATLAS-AFP~\cite{ATLAS-AFP}, where new forward proton detectors will be incorporated
(if funding is finally approved).
Although so far the statistics are very low these are expected
to be increased and become more relevant in future runs, as one is able to reach higher
energies~\cite{photon-scat-SM}.   

So far, LHC data have clearly indicated the presence of a relatively light Higgs-like
scalar  ($m_h\simeq 126$~GeV)
with coupling strengths close to the SM ones.
E.g. one has $a\approx 1$
for the $hWW$ coupling $a$ ($\kappa_V$ in other notations~\cite{ATLAS-exp,CMS-exp}),
within $\cO(10\%)$ uncertainties:
$a=1.15\pm 0.08$ (ATLAS~\cite{ATLAS-exp}), $a=1.01\pm 0.07$ (CMS~\cite{CMS-exp}).
Notice that a deviation  from its SM value $a=1$ implies immediately
a bad ultraviolet behaviour of the electroweak (EW) boson scattering amplitudes, which
needs of new beyond-SM (BSM) states to restore unitarity.
Thus, theoretical analyses have been performed on
the $WW$, $ZZ$ and $hh$ scattering amplitudes~\cite{WW-theory}
and ATLAS has recently provided the evidence of $W^\pm W^\pm$
scattering and the first rough bounds on the relevant low-energy ECLh parameters~\cite{WW-exp}.

In order to pin down these subtle effects one needs to go beyond using
simple effective vertices
to describe transitions such as $\gamma\gamma\to W^+ W^-$.
Instead one needs to properly incorporate the EW loops
(both UV-divergences and finite parts),
being the ECLh framework the most appropriate one in the current situation
where there seems to be a large mass gap and no new BSM particle has been found below the TeV.
This also means that in general one has to keep in the phenomenological analysis
all the counterterms allowed by the symmetry, unless one considers
particular BSM models.

For the construction of the low-energy EFT Lagrangian we will consider just the
observed particles, i.e., the SM content. More precisely, we will restrict our
calculation to the bosonic sector including the Higgs and  the EW gauge bosons $W^a$
and Would-Be-Goldstone Bosons (WBGB) $w^a$.
We will base  our EFT  on a custodial symmetry pattern,
where the spontaneous symmetry breaking (SSB)  $SU(2)_L\times SU(2)_R/SU(2)_{L+R}$
gives place to the WBGB. We will  not specify the origin of the SSB
and the WBGB will be non-linearly realized.
For simplicity, we perform our one-loop analysis in the Landau gauge,
where the ghost sector decouples and
the WBGBs become massless. In any case, in order to have a well defined classification
of the scaling of the diagrams one must work within renormalizable $R_\xi$  gauges,
which allow us to use the Equivalence Theorem (Eq.Th.)~\cite{equivalence-theorem}
and ensure an adequate $1/k^2$ scaling of the EW gauge boson propagators
in the UV~\cite{photon-scat,EW-chiral-counting}.
Thus, we will make use of the Eq.Th.~\cite{equivalence-theorem},
\bear
\mM(\gamma\gamma\to W_L^a W_L^b) &\simeq& \, -\, \mM(\gamma\gamma\to w^a w^b) \, ,
\eear
and extract the observable in the energy regime $m_{W}^2, m_{Z}^2\ll s$,
with $W^a$ and $w^a$ the EW gauge bosons and the WBGB, respectively.
The EW gauge boson masses $m_{W,Z}$ are then neglected
in our computation.~\footnote{
One must be careful with this limit, since $m_{W}^2 =g^2 v^2/4$, $m_Z^2=(g^2+g^{'2})v^2/4$
are set to zero while $v$
is kept fixed. This means we are taking $g,g'\to 0$  and considering just the leading
non-vanishing contribution to the photon scattering amplitude,
neglecting higher powers of $g^{(')}$.
}
Furthermore, since $m_h\sim m_{W,Z}\ll 4\pi v=3$~TeV
(with the vacuum expectation value $v=246$~GeV)
we also neglect the Higgs mass in our calculation, which means
not including Higgs mass corrections, typically proportional
to~$\cO(m_h^2/(16\pi^2 v^2))$~\cite{photon-scat,WW-theory}.
%
In summary, the applicability range of our analysis~\cite{photon-scat} is
\bear
m_W^2,\, m_Z^2 ,\, m_h^2 \quad \stackrel{\rm Eq.Th.}{\ll} \quad
s,\, t,\, u \quad \stackrel{\rm EFT}{\ll}  \quad \Lambda_{\rm ECLh}^2 \, ,
\eear
with the upper limit given by the  EFT cut-off  $\Lambda_{\rm ECLh}$,
expected to be of the order  of $4\pi v=3$~TeV or the mass of possible heavy BSM particles.
In principle, a more complete calculation beyond Eq.Th.
would allow to extend our ECLh result up to near-threshold energies
where, however, BSM effects are much more suppressed.

\vspace*{-0.25cm}
\section{The ECLh Lagrangian relevant for $\gamma\gamma\to w^aw^b$}

%
The ECLh  Lagrangian contains as dynamical fields the EW gauge bosons, $W^{\pm}$, $Z$ and $\gamma$,
the corresponding WBGBs ($w^{\pm}$ and $z$) and the Higgs-like scalar boson, $h$.
The WBGBs are described by a matrix field $U$ that takes values
in the $SU(2)_L \times SU(2)_R/SU(2)_{L+R}$ coset,   and transforms as $U \to L U R^\dagger$
under the action of the global group  $SU(2)_L \times SU(2)_R$
that defines the EW Chiral symmetry~\cite{Appelquist:1980vg,Longhitano:1980iz}.
The relevant ECLh Lagrangian with the basic building blocks are
\begin{eqnarray}
D_\mu U &=& \partial_\mu U + i\hat{W}_\mu U - i U\hat{B}_\mu,
\qquad
V_\mu = (D_\mu U) U^\dagger ,\;
\label{VmuandT}
\label{eq.cov-deriv} \\
\hat{W}_{\mu\nu} &=& \partial_\mu \hat{W}_\nu - \partial_\nu \hat{W}_\mu
 + i  [\hat{W}_\mu,\hat{W}_\nu ],
\quad 
\hat{B}_{\mu\nu}  = \partial_\mu \hat{B}_\nu -\partial_\nu \hat{B}_\mu ,
\label{fieldstrength}
\quad \hat{W}_\mu = g W_\mu^a  \tau^a/2 ,\;\hat{B}_\mu = g'\, B_\mu \tau^3/2 .
\label{EWfields}
\end{eqnarray}
Two particular parametrizations of the unitary matrix $U$ in terms of the dimensionless
$w^{\pm}/v$ and $z/v$ fields (exponential, with $U=\exp\{i \tau^a w^a/v\}$,
and spherical, with $U=\sqrt{1- w^a w^a/v^2}+ i\tau^a w^a/v$),
were considered in~\cite{photon-scat}, both leading to the
same predictions for the physical (on-shell) observables.~\footnote{
Other representations have been recently studied in Ref.~\cite{Machado:2014}.}
In order to have a power counting consistent with
the loop expansion we consider
 $\partial_\mu\, ,(g v)\, , (g'v), m_h \sim \cO(p)$ or, equivalently,
 $\partial_\mu\, ,m_W\, , m_Z, m_h \sim \cO(p) $,
and the scaling of the tensors
$D_\mu U,\; V_\mu\sim \cO(p)$ and
$\hat{W}_{\mu\nu},\;\hat{B}_{\mu\nu} \sim  \cO(p^2)$~\cite{photon-scat,EW-chiral-counting}.

With these building blocks one then constructs the ECLh up to a given order
in the chiral expansion. We require this Lagrangian to be CP invariant,
Lorentz invariant and $SU(2)_L \times U(1)_Y$ gauge invariant.
We focus ourselves on the relevant terms for $\gamma \gamma \to w^+ w^-$ and  $\gamma \gamma \to z z$ scattering processes. First, the relevant terms in the leading order (LO) Lagrangian --of $\cO(p^2)$-- are given by
\bear
\mL_2 &=&    -\Frac{1}{2 g^2} {\rm Tr}(\hat{W}_{\mu\nu}\hat{W}^{\mu\nu}) -\Frac{1}{2 g^{'2}} {\rm Tr} (\hat{B}_{\mu\nu} \hat{B}^{\mu\nu})
+\Frac{v^2}{4}\left[%
  1 + 2a \Frac{h}{v} + b \Frac{h^2}{v^2}\right] {\rm Tr} (D^\mu U^\dagger D_\mu U )
 + \Frac{1}{2} \partial^\mu h \, \partial_\mu h
 + \dots\, ,
 \nn\\
\label{eq.L2}
\eear
where the dots stand for $\cO(p^2)$ operators with three
or more  Higgs fields, which do not enter into this computation.
As already said, the Higgs mass term is neglected is our analysis~\cite{photon-scat}.
At NLO, we will have contributions from
the $O(p^4)$ Lagrangian~\cite{photon-scat,Longhitano:1980iz},
\bear
\mL_4 &=&
  a_1 {\rm Tr}(U \hat{B}_{\mu\nu} U^\dagger \hat{W}^{\mu\nu})
  + i a_2 {\rm Tr} (U \hat{B}_{\mu\nu} U^\dagger [V^\mu, V^\nu ])
  - i a_3  {\rm Tr} (\hat{W}_{\mu\nu}[V^\mu, V^\nu])
 -\Frac{c_{\gamma}}{2}\Frac{h}{v}e^2 A_{\mu\nu} A^{\mu\nu}\, +\, ...
\label{eq.L4}
\eear
involving the photon
field strength, $A_{\mu \nu}=\partial_\mu A_\nu- \partial_\nu A_\mu$.  The dots stand
for other $\cO(p^4)$ operators not contributing to $\gamma\gamma$ scattering.
Notice that the singlet Higgs field $h$ enters in the ECLh Lagrangian
via multiplicative polynomial functions  
and their derivatives.
%

\vspace*{-0.25cm}
\section{Analytical results for $\gamma\gamma\to w^aw^b$
in ECLh up to NLO}

In dimensional regularization, our NLO computation of  the $\mM(\gamma\gamma\to w^a w^b)$
amplitudes can be systematically  sorted out in the form~\cite{photon-scat}
\begin{equation}
\mM=\mM_{\rm LO}+\mM_{\rm NLO}
\quad \sim\quad
\underbrace{\cO(e^2)}_{\rm LO,\, tree}
\quad +\quad \left(
\underbrace{\cO\left( e^2\Frac{p^2}{16\pi^2 v^2}\right) }_{\rm NLO,\, 1-loop}
\quad+\quad
\underbrace{\cO\left( e^2 \Frac{a_i p^2}{v^2}\right) }_{\rm NLO,\, tree}
 \right) \,  .
 \label{eq.M}
\end{equation}
The LO contributions are given by all the tree-level diagrams
with vertices from just the $\mL_2$ Lagrangian.
The NLO contributions come from two types of graphs:
one-loop diagrams with only vertices  from $\mL_2$
(which may {\it a priori} generate  UV divergences);
tree-level diagrams with only one vertex  from  the $\mL_4$ Lagrangian, being
the remaining ones from  $\mL_2$.
The $a_i$ represent generic $\mL_4$ couplings ($a_1,\, a_2,\, a_3,\, c_\gamma$ in our case)
and will renormalize the referred one-loop UV divergences    
through an appropriate renormalization
${ a_i^r=a_i+\delta a_i }$ \cite{photon-scat,WW-theory,Machado:2014},
following a procedure completely analogous to that in Chiral Perturbation Theory~\cite{chpt}.
The $\delta a_i$ counterterms in turn, will lead to the corresponding running of the
renormalized ECLh parameters,
i.e. the dependence with the renormalization scale $\mu$ of the  $a_i^r(\mu)$.

The amplitudes $\mM(\gamma(k_1,\epsilon_1)\gamma(k_2,\epsilon_2)\to w^a(p_1) w^b(p_2))$
with $w^a w^b=zz,w^+w^-$, are given in terms of the Lorentz
decomposition~\cite{photon-scat}
\be
\mM   =
ie^2 (\epsilon_1^\mu \epsilon_2^\nu T_{\mu\nu}^{(1)}) A(s,t,u)+
ie^2 (\epsilon_1^\mu \epsilon_2^\nu T_{\mu\nu}^{(2)})B(s,t,u),
\eequ
written in terms of the two independent Lorentz structures
involving the external momenta, which can be found in~\cite{photon-scat}.
The Mandelstam variables are defined as
$s=(p_1+p_2)^2$, $t=(k_1-p_1)^2$ and $u=(k_1-p_2)^2$ and the $\epsilon_i$'s
are the  polarization vectors of the external photons.

For the scalar functions $A$ and $B$ we will follow the same notation
as in Eq.~(\ref{eq.M}): $A=A_{\rm LO}+A_{\rm NLO}$,
$B=B_{\rm LO}+B_{\rm NLO}$.  Thus,
At LO one has the well-known tree-level contribution from $\mL_2$,
\bear
A(\gamma\gamma\to zz)_{\rm LO}
&=& B(\gamma\gamma\to zz)_{\rm  LO}
= 0 ,
\\
A(\gamma\gamma\to w^+ w^-)_{\rm LO}    
&=& 2 s B(\gamma\gamma\to w^+ w^-)_{\rm  LO}    
= -\Frac{1}{t} - \Frac{1}{u}.
\eear
At NLO, i.e at  $\cO(e^2p^2)$, we find
\bear
A(\gamma \gamma \to zz)_{\rm NLO}&=&
\Frac{2 a c_\gamma^r}{v^2} + \Frac{ (a^2-1)}{4\pi^2v^2}\, ,
\label{eq.A-zz}
\qquad\qquad\qquad  \qquad\qquad
B(\gamma \gamma \to zz)_{\rm NLO}= 0,
\label{eq.B-zz}
\\
A(\gamma \gamma \to w^+w^-)_{\rm NLO}&=&
\Frac{2 a c_\gamma^r}{v^2} + \Frac{ (a^2-1)}{8\pi^2v^2}
+ \Frac{8(a_1^r-a_2^r+a_3^r)}{v^2}\, ,
\label{eq.A-zz}
\qquad
B(\gamma \gamma \to w^+w^-)_{\rm NLO}= 0.
\label{eq.results}
\eear
The term with $c_\gamma^r$  comes from the Higgs tree-level
exchange in the $s$--channel, the term proportional to $(a^2-1)$
comes from the one-loop diagrams with $\mL_2$ vertices,
and the Higgsless operators in Eq.~(\ref{eq.L4}) yield the tree-level contribution
to $\gamma\gamma\to w^+ w^-$ proportional to $(a_1-a_2+a_3)$.
It is worthy to remark that the one-loop amplitude does not carry the typical
chiral suppression $\cO(E^2/(16\pi^2 v^2))$~\cite{Morales:92}
(with $E^2=s,t,u$) but
a stronger one $\cO((a^2-1) E^2/(16 \pi^2 v^2))$.  Notice that the
experimental determinations of $a$ find it to be
close to 1 within an $\cO(10\%)$ uncertainty~\cite{ATLAS-exp,CMS-exp}.
Thus, in the limit $a\to 1$ the loop contribution vanishes, as one finds in the SM
under the same approximations consider in our work~\cite{photon-scat}.

Independent diagrams are  in general UV divergent and have complicated
Lorentz structures and energy dependence (see App.~B in~\cite{photon-scat} for details).
However, in dimensional regularization,
the final one-loop amplitude turns out to be UV finite
after all the contributions are put together.
The reason for this subtle cancellation can be understood from the equivalence between
our loop diagrams and the one-loop amplitude in the
$SO(5)/SO(4)$ Non-Linear Sigma Model~\cite{photon-scat,MCHM}
in the approximations considered in our analysis~\cite{photon-scat}.
Thus, the combinations of
$\mL_4$ chiral parameters $(a_1^r-a_2^r+a_3^r)$ and
$c^r_\gamma$ do not need to be renormalized:
\begin{equation}
a_1^r-a_2^r+a_3^r=a_1 -a_2 +a_3 \, ,\qquad c^r_\gamma=c_\gamma.
\label{renormcgamma}
\end{equation}
Notice that when setting $a=c_\gamma =0$ in our formulas above
we recover exactly the previous result found in~\cite{Morales:92}
for the case of the Higgsless EW Chiral Lagrangian (ECL)  and the analogous
result for the amplitude in the pion case~\cite{photon-scat-ChPT}.

\section{Related observables, global analysis and running determination}

As we can see in Eq.~(\ref{eq.results})  the $\gamma\gamma\to zz$ and $\gamma\gamma\to w^+w^-$
cross sections depend on three independent combinations of parameters: $a$, $c_\gamma$
and $(a_1-a_2+a_3)$. In order to pin down each of these ECLh couplings one must
combine our $\gamma\gamma$-scattering analysis with other observables that depend
on this same set of  parameters. It is not difficult to
find that other processes involving photons depend on these parameters.
In Ref.~\cite{photon-scat} we computed 4 more observables of this kind:
the $h\to \gamma\gamma$ decay width, the oblique $S$--parameter, and the
$\gamma^*\to w^+w^-$ and $\gamma^*\gamma \to h$ electromagnetic form-factors.
In Table~\ref{tab.relevant-couplings} we provide the relevant combinations of
$\mL_2$ and $\mL_4$  coefficients in each case. The detailed amplitudes
can be found in App.~D in~\cite{photon-scat}.
These six observables allow the extraction of the four independent combinations of couplings
$a$,$c_\gamma$, $a_1$ and $(a_2-a_3)$. Moreover, one can use four of these observables
to perform a prediction for the remaining two.

\begin{table}[!t]
\begin{center}
\begin{tabular}{l|cc}
{\bf Observables}     & \multicolumn{2}{c}{\bf Relevant combinations of parameters} \\
\null  & \ \ from $\mL_2$\ \  & from $\mL_4$ \\
\hline
$\mM(\gamma\gamma\to zz)$     & $a$ & $c_\gamma^{r}$ \\
$\mM(\gamma\gamma\to w^+w^-)$ & $a$ & $(a_1^r- a_2^r+a_3^r),\, c_\gamma^{r}$ \\
$\Gamma(h\to\gamma\gamma)$    & $a$ & $c_\gamma^{r}$ \\
$S$--parameter                & $a$ &
    $a_1^r$
\\
$\mF_{\gamma^*ww}$            & $a$ &
      $(a_2^r-a_3^r)$
\\
$\mF_{\gamma^*\gamma h}$      & --  & $c_\gamma^{r}$\\
\hline  
\end{tabular}
\caption{{\small
Set of six observables studied in Ref.~\cite{photon-scat} and their corresponding  relevant
combinations of chiral parameters.
}}
\vspace*{-0.5cm}
\label{tab.relevant-couplings}
\end{center}
\end{table}

The one-loop contribution in the six relevant amplitudes is found to
be UV-divergent in some cases. These divergences are absorbed by means
of the renormalizations $a_i^r(\mu)=a_i+\delta a_i$ and the $\cO(p^4)$
counterterms $\delta a_i$
(with $a_i$ referring to the corresponding $\mL_4$ coefficients
$a_i=c_\gamma$, $a_1$, $a_2$, $a_3$).
As expected, the renormalization in the six observables gives a fully consistent
set of renormalization conditions
and fixes the running of the renormalized couplings in the way
\bear
\Frac{da_i^r}{d\ln\mu} \,=\, -\, \Frac{\Gamma_{a_i}}{16\pi^2 }\, ,
\eear
with the corresponding $\Gamma_{a_i}$ given in Table~\ref{tab.running}.
For the sake of completeness,
we have added the running of the ECLh parameters $a_4^r$ and $a_5^r$,
which have been recently determined in the one-loop analysis of  $WW$--scattering
within the framework of chiral Lagrangians~\cite{WW-theory}.
One can see that in the SM limit $a=b=1$ the $\mL_4$ coefficients
in Table~\ref{tab.running}  do not run,
in agreement with the fact that these higher order operators are absent in the SM.

\begin{table}[!t]
\begin{center}
\begin{tabular}{ c|c|c }
\rule{0pt}{3ex}
 & {\bf ECLh  } &  {\bf ECL} (Higgsless)                 
\\[5pt] \hline
\rule{0pt}{3ex}
$\quad \Gamma_{a_1-a_2+a_3}\quad$ & $\quad 0 \quad $ &  0                
\\[5pt] \hline
\rule{0pt}{3ex}
$\quad \Gamma_{c_\gamma}\quad$ & $\quad 0 \quad $ &  -                
\\[5pt] \hline
\rule{0pt}{3ex}
$\quad \Gamma_{a_1}\quad$ & $\quad -\frac{1}{6}(1-a^2) \quad $&$\quad -\frac{1}{6} \quad $                
\\[5pt] \hline
\rule{0pt}{3ex}
$\quad \Gamma_{a_2-a_3} \quad$ & $\quad -\frac{1}{6}(1-a^2) \quad $& $\quad -\, \frac{1}{6} \quad $                  
\\[5pt] \hline
\rule{0pt}{3ex}
$\quad \Gamma_{a_4}\quad$ & $\quad \frac{1}{6}(1-a^2)^2  \quad $ & $\quad \frac{1}{6}\quad $                 
\\[5pt] \hline
\rule{0pt}{3ex}
$\quad \Gamma_{a_5}\quad$ & $\quad \frac{1}{8}(b-a^2)^2
+\frac{1}{12}(1-a^2)^2\quad $ &
$\quad \frac{1}{12}\quad $                 
\\[5pt] \hline
\end{tabular}
\caption{\small
Running of the relevant ECLh parameters and their combinations appearing
in the six selected observables.
The third column provides the corresponding running
for the Higgsless ECL case~\cite{Morales:94}.
}
\vspace*{-0.5cm}
\label{tab.running}
\end{center}
\end{table}


To end up the discussion we would like to point out a series of improvements
that will be incorporated to the $\gamma\gamma$ analysis~\cite{photon-scat}
in a future phenomenological analysis.
First, one must take into account fermion loops and in particular the top quark contribution,
which is expected to be as numerically important as the $W$ loops, as it happens in analogous processes
like e.g. $h\to\gamma\gamma$.
Likewise, the computation can be refined by taking into account the impact
of the $W$, $Z$ and $h$ masses, allowing the extension of the analysis all the way
down to the production threshold. Thus, one should go beyond the Equivalence Theorem
and compute the actual $\gamma\gamma\to W_L^a W_L^b$ amplitudes. These light mass effects
are nevertheless suppressed below the percent level for energies of the order of the TeV.

We plan to perform a MonteCarlo analysis for LHC and future $e^+ e^-$ accelerators
in order to estimate both integrated even rates and the energy dependence.
Previous studies expect a larger relevance of the
$\gamma\gamma\to W^+W^-$ subprocess  in the $W^+W^-$ production
at future LHC runs, which can even exceed the $gg\to W^+W^-$ contribution to the cross section
at large $p_T$~\cite{photon-scat-SM}.


\end{document}